\documentclass{article}
\usepackage{lajolla2006}
\usepackage{graphicx,amsfonts,mathrsfs,amssymb,amsmath,amsopn,bm,mathbbol,qmech}

\def\rightmark{Medium Modifications of Vector Mesons and NA60}
\def\leftmark{Hendrik van Hees et al.}
\frompage{000} \topage{000}

\title{Medium Modifications of Vector Mesons and NA60} 

\authors{
{Hendrik van Hees and Ralf Rapp$^1$ %
}\\[2.812mm]
{\normalsize
\hspace*{-8pt}$^1$Cyclotron Institute and Physics Department, Texas A{\&}M
  University, College Station, Texas 77843-3366, USA 
}}

\date{March 13, 2006} 
\keyword{ultra-relativistic heavy-ion collisions, dileptons, chiral symmetry} 
\PACS{25.75.-q,12.40.Vv,13.75.-n,11.40.-q}

\abstract{We confront different models for medium modifications of the
  electromagnetic correlation function to recent dimuon spectra from the
  NA60 collaboration for central In-In collisions at the CERN-SPS. In
  the low-mass region ($M$$\lesssim$$1.0 \; \text{GeV}$), we evaluate
  medium modifications of the spectral properties of $\rho$-, $\omega$-
  and $\phi$-mesons. In the intermediate-mass region
  ($M$$>$$1.0\;\text{GeV}$), we employ schematic models for the mixing
  of vector and axialvector current correlators, which provide a
  mechanism for chiral-symmetry restoration.}

\begin{document}
\maketitle
\setcounter{page}{1}

%%%%%%%%%%%%%%%%%%%%%%%%%%%%%%%%%%%%%%%
\section{Introduction}
%%%%%%%%%%%%%%%%%%%%%%%%%%%%%%%%%%%%%%%

One of the main goals in the investigation of hot and dense matter
produced in ultra-relativistic heavy-ion collisions (URHICs) is the
understanding of the phase structure of Quantum Chromodynamics (QCD),
the theory of the strong interaction. At low temperatures the relevant
degrees of freedom are hadrons, and the approximate chiral symmetry of
the light-quark sector is spontaneously broken. From asymptotic freedom
of QCD one expects that at high temperatures and/or densities the
partons are deconfined. Lattice-QCD simulations show a phase transition
at a critical temperature of about $T_{c}$$\simeq$$175\;\text{MeV}$,
above which chiral symmetry is restored and the partons are
deconfined~\cite{KL03}. Since chiral-symmetry restoration implies that
chiral multiplets in the hadron spectrum degenerate, significant
in-medium modifications of hadron properties in the hadronic phase are
expected.

Electromagnetic (e.m.) probes, i.e., leptons and photons, are
particularly valuable for the investigation of the hot and dense medium
since they are produced during all stages of the collision and leave the
fireball nearly undisturbed by final-state interactions. An important
finding in the CERN Super Proton Synchrotron (SPS) program in the 1990's
is the large excess of low-mass dileptons ($M$$\lesssim$$1\;\text{GeV}$)
in Pb-Au collisions compared to expectations from proton-proton
collisions~\cite{ceres05}. This result has been explained by medium
modifications of the $\rho$ meson~\cite{rw99}, but no decisive
distinction between a broadening and a dropping mass could be made.

The objective of this talk is to evaluate different approaches to
low-mass vector mesons in a hot and dense medium relative to recent
dimuon spectra in In-In collisions by the NA60 collaboration at the
SPS~\cite{dam05}. For the first time a discrimination of different
theoretical mechanisms for chiral-symmetry restoration has become
possible. We also investigate the mass-region beyond the $\phi$ ($1
\;\text{GeV}$$<$$M$$<$$1.5 \; \text{GeV}$) within schematic models which
implement a mixing of the vector and the axialvector current correlators,
leading to their degeneracy at the critical point as a consequence of
chiral-symmetry restoration.

%%%%%%%%%%%%%%%%%%%%%%%%%%%%%%%%%%%%%
\section{Dilepton Rates and Spectra}
%%%%%%%%%%%%%%%%%%%%%%%%%%%%%%%%%%%%%

The differential rate for the production of lepton pairs from a thermal
source can be expressed in terms of the retarded e.m. current
correlator, $\Pi_{R}^{\mu \nu}$~\cite{MT84,wel90,gale-kap90},
\begin{equation}
\label{del-rate}
\frac{\d N_{ll}}{\d^4 x \d^4 q}=\frac{\d R}{\d^4 q}=-\frac{\alpha^2}{3
  \pi^3} \frac{L(M^2)}{M^2} \im \Pi_{R \mu}^{\mu}(q) f^B(q_0),
\end{equation}
where $L(M)$ is the leptonic phase-space integral and $f_B$ the Bose
distribution.

Dilepton spectra in URHICs are obtained upon convoluting the above rate
over the space-time evolution of the fireball,
\begin{equation}
\frac{\d N_{ll}}{\d M}=\int_0^{t_{\rm fo}} \d t V_{\text{FB}}(t) \int
\frac{\d^3 q}{q_0} \frac{\d N_{ll}}{\d^4 x \d^4 q} z_{\pi}^n
\frac{M}{\Delta y} A(M,q_t,y) \ ,
\end{equation}
where an average over the rapidity window, $\Delta y$, and the
integration over the three-momentum of the pair have been carried out.
$A$ is the detector acceptance function which we have determined from
NA60 simulations~\cite{dam06}. The pion-fugacity factor,
$z_{\pi}^n$$=$$\exp(n \mu_{\pi}/T)$, accounts for chemical
off-equilibrium in the hadronic phase, with $n$$=$$2,3,4$ for $\rho$,
$\omega$, and four-pion contributions to the dilepton rates,
respectively.  The thermal-fireball volume is assumed to expand
cylindrically (and isotropically)~\cite{rw99b},
$V_{\text{FB}}(t)$$=$$(z_0+v_z t)\pi (r_{\perp}+0.5 a_{\perp}t^2)^2$.
For central In(158~$A$GeV)-In we use a transverse acceleration
$a_{\perp}$=$0.08\,c^2/\text{fm}$, longitudinal speed $v_z$$=$$c$, and
initial sizes $z_0$=1.8~fm/$c$, $r_{\perp}$=5.15~fm. With a
hadron-resonance gas equation of state and chemical freezeout at
$(\mu_{N}^{\text{ch}},T_{\text{ch}})$$=$$(232,175) \;\text{MeV}$, a
total fireball entropy of $S$$=$$2630$ translates into $\d
N_{\text{ch}}/\d y$$\simeq$$195$. Assuming isentropic expansion we infer
the temperature and baryon density from the entropy density,
$s(t)$=$S/V(t)$. The initial QGP temperature is $T_0$$=$$197\;
\text{MeV}$, at $T_{\text{ch}}$$\simeq$$T_c$ the system converts into
hadronic matter, and the time evolution terminates with thermal
freezeout at $t_{\rm fo}$=7~fm/$c$ ($T_{\text{fo}}$$\simeq$$120 \;
\text{MeV}$).

%%%%%%%%%%%%%%%%%%%%%%%%%%%%%%%%%%%%%%%%%%%%%%%%%%%%%
\section{Hadronic Many-Body Approach and ``Duality''}
%%%%%%%%%%%%%%%%%%%%%%%%%%%%%%%%%%%%%%%%%%%%%%%%%%%%%

In this Section the e.m.~correlator, $\Pi_{R}^{\mu \nu}$, will be
described by a combination of hadronic many-body theory (HMBT) for
$\rho$, $\omega$ and $\phi$ at low mass, by lowest-order-in-$T$ chiral
mixing at intermediate mass, and by a hard-thermal loop improved
emission in the QGP~\cite{bpy90}.  The vector-meson propagators are
directly related to the e.m.~correlator employing the vector-dominance
model,
\begin{equation}
\label{vdm}
\im \Pi_{R\mu}^{\mu} = \sum_{V=\rho,\omega,\phi}
\frac{1}{g_V^2} \left(m_{V}^{(0)}
  \right)^4 \im D_{RV \mu}^{\mu},
\end{equation}
($g_V$, $m_{V}^{(0)}$: vector-meson coupling constants and their (bare)
masses).

A particular appealing feature that emerges in this approach is the
approximate degeneracy of the in-medium hadronic rate with the QGP one
close to $T_c$, which has been interpreted has a reduction of the
quark-hadron duality scale in the medium~\cite{Rapp:qm99} and provides a
"natural" scenario for chiral restoration.

%%%%%%%%%%%%%%%%%%%%%%%%%%%%%%%%%%%%%%%
\subsection{$\rho$-Meson in Medium}
%%%%%%%%%%%%%%%%%%%%%%%%%%%%%%%%%%%%%%%

The dominant contribution to Eq.~(\ref{vdm}) is the isovector part which
at low mass is saturated by the $\rho$-meson. We here use the in-medium
$\rho$ spectral function as evaluated in hadronic many-body theory in
Ref.~\cite{rw99b}. Starting point is a realistic description of the
(pion cloud of the) $\rho$ in the vacuum, consistent with $P$-wave
$\pi$-$\pi$ scattering phase shifts and the pion e.m.~form factor as
encoded in the one-loop self-energy.

In hadronic matter, the pion cloud is dressed by Bose enhancement
factors and pion-induced $NN^{-1}$ and $\Delta N^{-1}$ excitations
("pisobars").  To approximately account for higher nucleon and thermally
excited baryon resonances ($B^*$), an effective baryon density,
$\rho_{\text{eff}}$$=$$\rho_N$$+$$\rho_{B^*}/2$, has been applied. In
addition, direct $\rho$-$BB^{-1}$ excitations ("rhosobars") on nucleons,
$\Lambda$'s, etc., are included. In cold nuclear matter, the interaction
vertices (coupling constants and form factors) have been constrained by
a comprehensive fit to photo-absorption on the nucleon and
nuclei~\cite{rubw98} (including an extended vector dominance
model~\cite{klz67}), as well as by $\pi N\to \rho N$ scattering.

Meson-gas contributions to the self-energy are calculated following
Ref.~\cite{gale-rapp99} including a rather complete set of $s$-channel
resonances up to $1.65$~GeV with interaction vertices constrained by
hadronic and radiative decay branchings.

As a result the in-medium $\rho$ spectral function exhibits substantial
broadening together with a small upward mass shift. This behavior is
typical for hadronic many-body calculations since contributions to the
imaginary part of the retarded self-energy are strictly of the same sign
due to the retardation condition, while the real parts contain both
attractive (negative) and repulsive (positive) interactions, which for a
large set of excitations tend to compensate each other.

%%%%%%%%%%%%%%%%%%%%%%%%%%%%%%%%%%%%%%%%%%%%%%%%%%
\subsection{$\omega$- and $\phi$-Meson in Medium}
%%%%%%%%%%%%%%%%%%%%%%%%%%%%%%%%%%%%%%%%%%%%%%%%%%

The $\omega$-meson is treated along similar lines as the
$\rho$-meson~\cite{Rapp01}: the vacuum self-energy is constructed from a
combination of $\rho\pi$ and 3$\pi$ loops, whose couplings and form
factors have been adjusted to the radiative width (assuming vector-meson
dominance) and total hadronic decay width (with 50\% from $\rho\pi$ and
direct 3$\pi$ pieces each). Medium modifications in the $\pi \rho$ loop
are introduced by Bose-enhancement factors and the in-medium $\rho$
spectral function as described above. Meson-gas modifications include
the inelastic channel $\omega \pi$$\rightarrow$$\pi \pi$~\cite{hag95}
and scattering off thermal pions via the $b_1(1235)$ which is the
empirically dominant $\omega \pi$ resonance. The effects of direct
resonances on baryons have been assigned to $N(1520)N^{-1}$ and $N(1650)
N^{-1}$ excitations. The total in-medium width of the $\omega$ in
hadronic matter representative for URHICs amounts to
$\Gamma_{\omega}^{\text{med}}$$\simeq$100~MeV.

Recent developments render the situation for the $\phi$-meson more
involved. It's collisional broadening in a meson gas has been predicted
as large as $\Delta\Gamma_\phi^{\rm coll}$$\simeq$20-50~MeV at
$T$=150-180~MeV~\cite{AK02}. In cold nuclear matter, at saturation
density $\rho_0$$=$$0.16\;\text{fm}^{-3}$, dressing the kaon-cloud was
found to induce $\Delta\Gamma_\phi^{K\bar
  K}$$\simeq$25~MeV~\cite{CROTV04}. This result appears to underestimate
the absorption inferred from recent nuclear photoproduction data by
about a factor of 2~\cite{ahn05}. We therefore assume an in-medium
broadening which, when averaged over the fireball evolution, amounts to
$\Gamma_{\phi}^{\text{med}}$$\simeq$$80\;\text{MeV}$. Furthermore, the
$\phi$ abundance is corrected for a two-kaon fugacity factor, $z_K^2$,
and a strangeness-suppression factor, $\gamma_s^2$ with
$\gamma_s$$\simeq$$0.75$~\cite{BGKMS04}.

%%%%%%%%%%%%%%%%%%%%%%%%%%%%%%%%%%%%%%%%%%
\subsection{Four-Pion Contributions}
\label{sec_4pi}
%%%%%%%%%%%%%%%%%%%%%%%%%%%%%%%%%%%%%%%%%%

\begin{figure}[!t]
\begin{minipage}{0.48\textwidth}
\includegraphics[width=\textwidth]{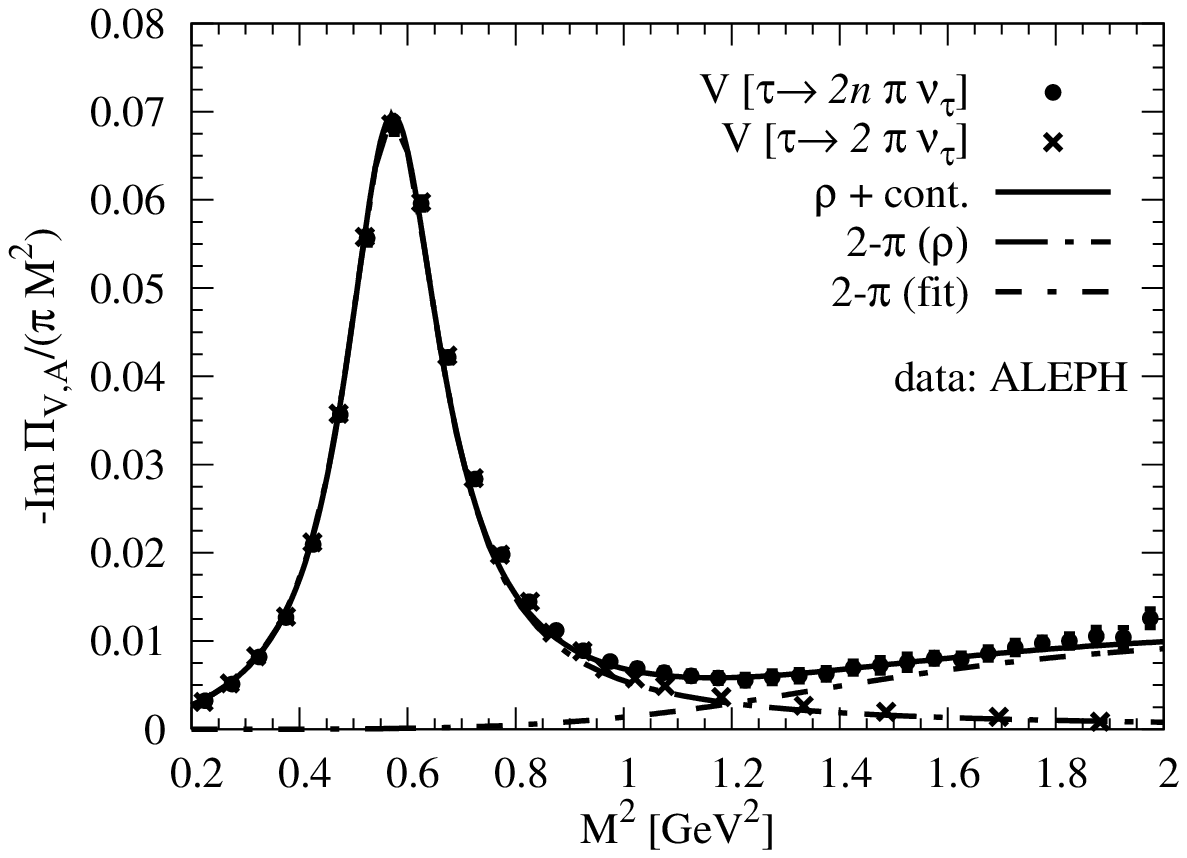}
\end{minipage}\hfill
\begin{minipage}{0.48\textwidth}
\includegraphics[width=\textwidth]{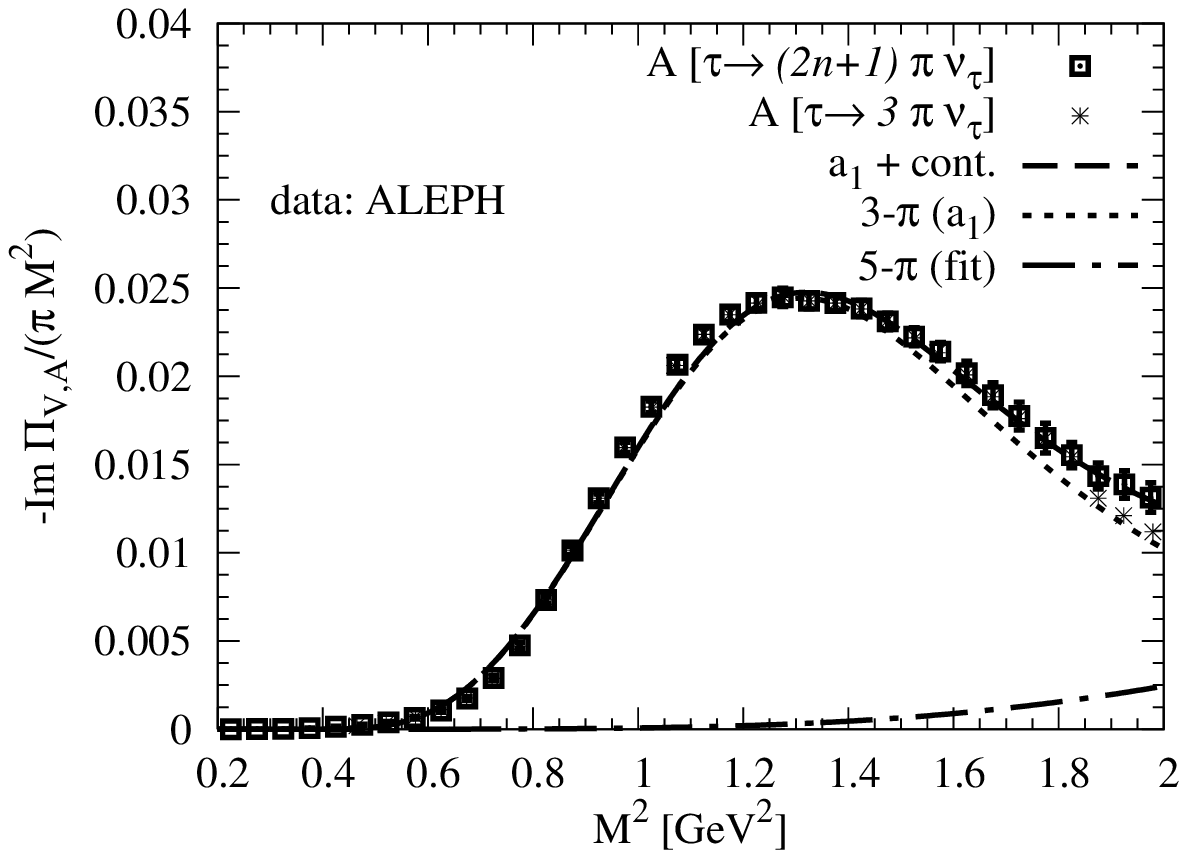}
\end{minipage}
\caption{\label{fig.1} Free isovector-vector (left panel) and
  -axialvector (right panel) current correlators from $\tau$-decay data,
  compared to fits to three- and four-pion contributions (the two-pion
  part corresponds to the vacuum $\rho$-spectral function).}
\end{figure}
At masses above 1~GeV, four-pion (and higher) states dominate the free
e.m.~correlator, cf.~left panel of Fig.~\ref{fig.1}. We implement medium
effects in this regime by employing model-independent predictions by
chiral symmetry. To lowest order in $T$ these amount to a pion-induced
mixing of vector ($V$) and axialvector ($A$) correlators~\cite{DEI90}:
\begin{equation}
\label{chir-mix}
\Pi_{V,A}=(1-\epsilon)~\Pi_{V,A}^{(\text{vac})}
+\epsilon~\Pi_{A,V}^{(\text{vac})},
\end{equation}
where $\epsilon$ is the mixing parameter ($\epsilon$=$T^2/6f_\pi^2$ in
the chiral limit, $f_\pi=93\;\text{MeV}$) arising from pion tadpole diagrams.
The admixture of the $a_1$ resonance (right panel of Fig.~\ref{fig.1})
entails an enhancement of the dilepton rate for $M$$\simeq$1-1.4~GeV.
We evaluate the tadpole integral including both finite pion-mass and
chemical potential~\cite{HR06a},
\begin{equation}
\label{thermal-tad}
I_{\pi}(T,\mu_{\pi})=\int \frac{\d^3 k}{(2 \pi)^3}
\frac{f^{\pi}[\omega_{\pi}(k),\mu_{\pi}]}{\omega_{\pi}(k)} 
\end{equation}
($\omega_{\pi}(k)$$=$$\sqrt{k^2+m_{\pi}^2}$). An upper estimate of the
mixing is then obtained by assuming $V$-$A$ degeneracy at $T_c$ by
setting
\begin{equation}
\label{max-mix}
\epsilon=\frac{1}{2} \frac{I_{\pi}(T,\mu_{\pi})}{I_{\pi}(T_c,\mu_{\pi}=0)}.
\end{equation}
The vacuum $V$ and $A$ spectral functions are fitted to $\tau$-decay
data from the ALEPH collaboration~\cite{aleph98} (Fig.~\ref{fig.1}).
Note that the two-pion contributions are omitted here as they are
included in the (in-medium) $\rho$-meson (which also incorporates the
mixing).

%%%%%%%%%%%%%%%%%%%%%%%%%%%%%%%%%%%%%%%%%%%%%%%
\subsection{Comparison to NA60 Data}
%%%%%%%%%%%%%%%%%%%%%%%%%%%%%%%%%%%%%%%%%%%%%%%
\begin{figure}[!t]
\begin{minipage}{0.48\textwidth}
\includegraphics[width=\textwidth]{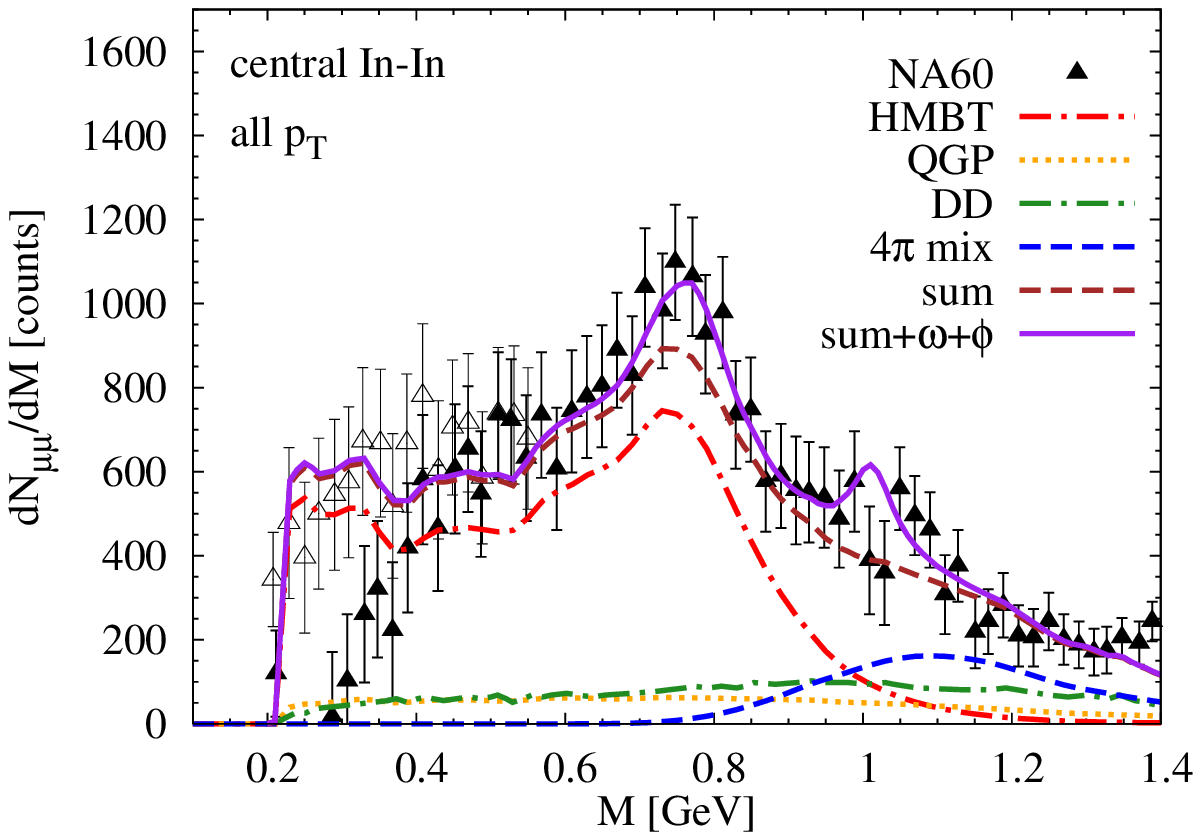}
\end{minipage}\hfill
\begin{minipage}{0.48\textwidth}
\includegraphics[width=\textwidth]{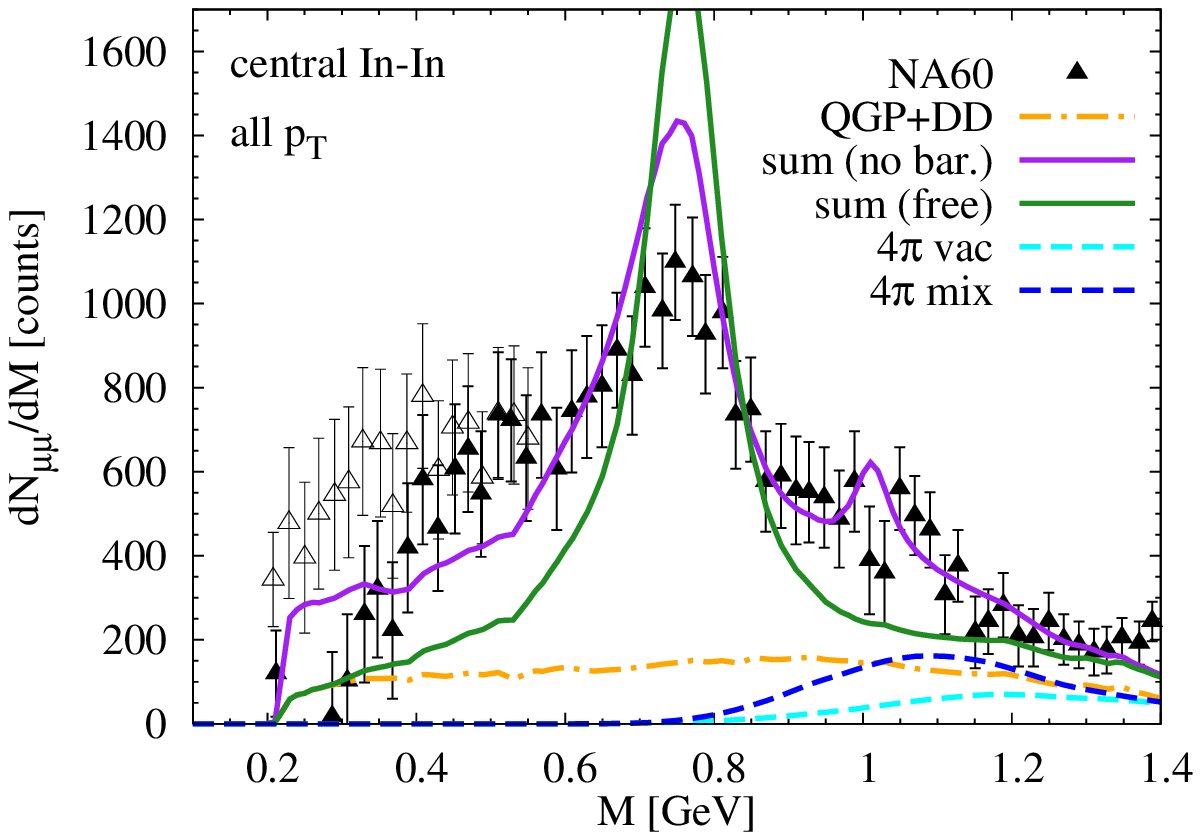}
\end{minipage}
\caption{\label{fig.2} NA60 data~\cite{dam05} compared to thermal
  $\mu^+\mu^-$ spectra from an expanding fireball using in-medium
  dilepton rates.  Left panel: hadronic many-body theory for vector
  mesons and chiral mixing for 4-$\pi$ contributions; right panel:
  removing the effects of baryons (``no bar.'') or neglecting medium
  effects altogether (``free'').}
\end{figure}
In Fig.~\ref{fig.2} (left panel) the combined thermal $\mu^+\mu^-$ yield
($\rho$, $\omega$, $\phi$, 4$\pi$ with mixing and QGP, convoluted over
the fireball) is compared to NA60 data in central In-In. The spectra are
well described with absolute normalization. Compared to earlier
predictions~\cite{rap04,dam05}, the fireball acceleration has been
increased which reduces the lifetime from 10 to 7~fm/$c$ and generates
harder $q_t$ spectra. We emphasize that the relative strength of the
various thermal sources is fixed, which renders the overall agreement
very encouraging. In the right panel of Fig.~\ref{fig.2} we illustrate
the importance of baryonic (medium) effects on the $\rho$, as well as
the enhancement due to chiral $V$-$A$ mixing. A reduction of the 
uncertainties in the $\eta$-cocktail subtraction (illustrated by the 
two data sets close to threshold) can further test these predictions.

%%%%%%%%%%%%%%%%%%%%%%%%%%%%%%%%%%%%%%%%%%%%%%%%%%%%%%%
\section{Alternative Approaches for Medium Modifications}
%%%%%%%%%%%%%%%%%%%%%%%%%%%%%%%%%%%%%%%%%%%%%%%%%%%%%%%

To further scrutinize the sensitivity of the NA60 data we here study the
consequences of different approaches to implement medium effects on the
e.m.~correlator.
 
%%%%%%%%%%%%%%%%%%%%%%%%%%%%%%%%%%%%%%%%%%%%%
\subsection{Chiral Reduction Formalism}
%%%%%%%%%%%%%%%%%%%%%%%%%%%%%%%%%%%%%%%%%%%%%

A model-independent treatment of the in-medium e.m.~correlator is
obtained by coupling a low-temperature and -density expansion with
chiral reduction techniques~\cite{syz97}, similar in spirit to
Sec.~\ref{sec_4pi} but in a more sophisticated way using vacuum
scattering amplitudes off pions and nucleons.  The amplitudes are
constructed from $\tau$-decay data (Fig.~\ref{fig.1}) and
photoabsorption on the nucleon. Appropriate fugacity factors,
$z_{\pi}^n$, are included according to the fireball as used above (for
simplicity, we neglect $\omega$ and $\phi$ contributions, but include
QGP radiation as before).

The left panel of Fig.~\ref{fig.3} compares pertinent thermal spectra to
NA60 data. The mass region above 1~GeV is well described, while in the
low-mass region the e.m.~spectral function produces a too narrow peak
structure around the $\rho$-mass. This is due to the density expansion
in the chiral reduction formalism which does not generate a broadening
of the $\rho$ spectral function.

%%%%%%%%%%%%%%%%%%%%%%%%%%%%%%%%%%%%%%%%%%%%%%
\subsection{A Dropping $\rho$-Mass Scenario}
%%%%%%%%%%%%%%%%%%%%%%%%%%%%%%%%%%%%%%%%%%%%%%

In the 1990's, a dropping $\rho$-mass has successfully been implemented
to describe the low-mass dilepton enhancement observed by
CERES/NA45~\cite{ceres05}. We here check the consequences of this
assumption for NA60 using an in-medium mass parameterization~\cite{rw99}
of type $m_{\rho}^*=m_{\rho} (1-C\rho_B/\rho_0)[1-(T/T_c)^2]^{\alpha}$,
where the density dependence with $C=0.15$ resembles QCD sum rule
estimates~\cite{Hatsuda}, while $\alpha$=0.3 is motivated by the
temperature evolution of the chiral condensate~\cite{Pelaez02}. The
pertinent $\rho$ spectral function (which has also been supplemented by
a small thermal broadening) is then evolved through the same fireball as
above, maintaining the absolute normalization of the spectra. The
$\rho$ contribution is furthermore supplemented with QGP and four-pion
components (including chiral mixing).

\begin{figure}[!t]
\begin{minipage}{0.48\textwidth}
\includegraphics[width=\textwidth]{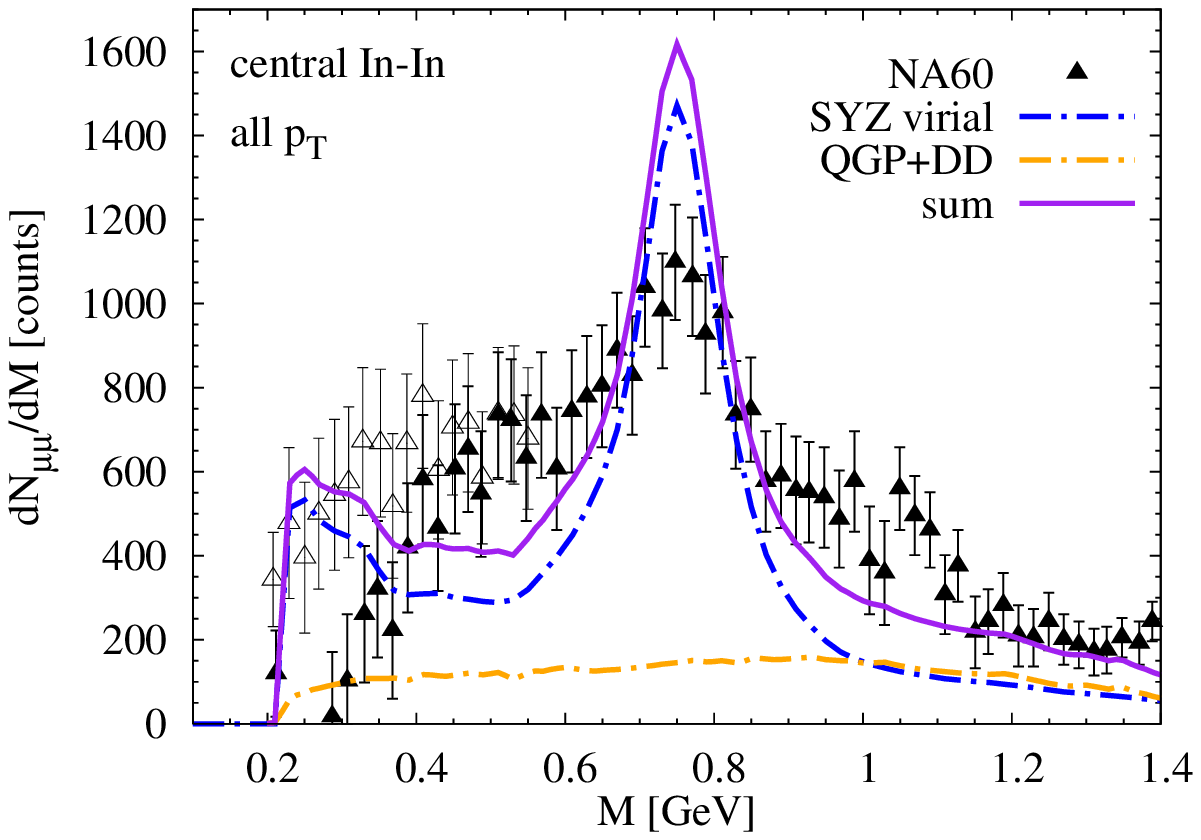}
\end{minipage}\hfill
\begin{minipage}{0.48\textwidth}
\includegraphics[width=\textwidth]{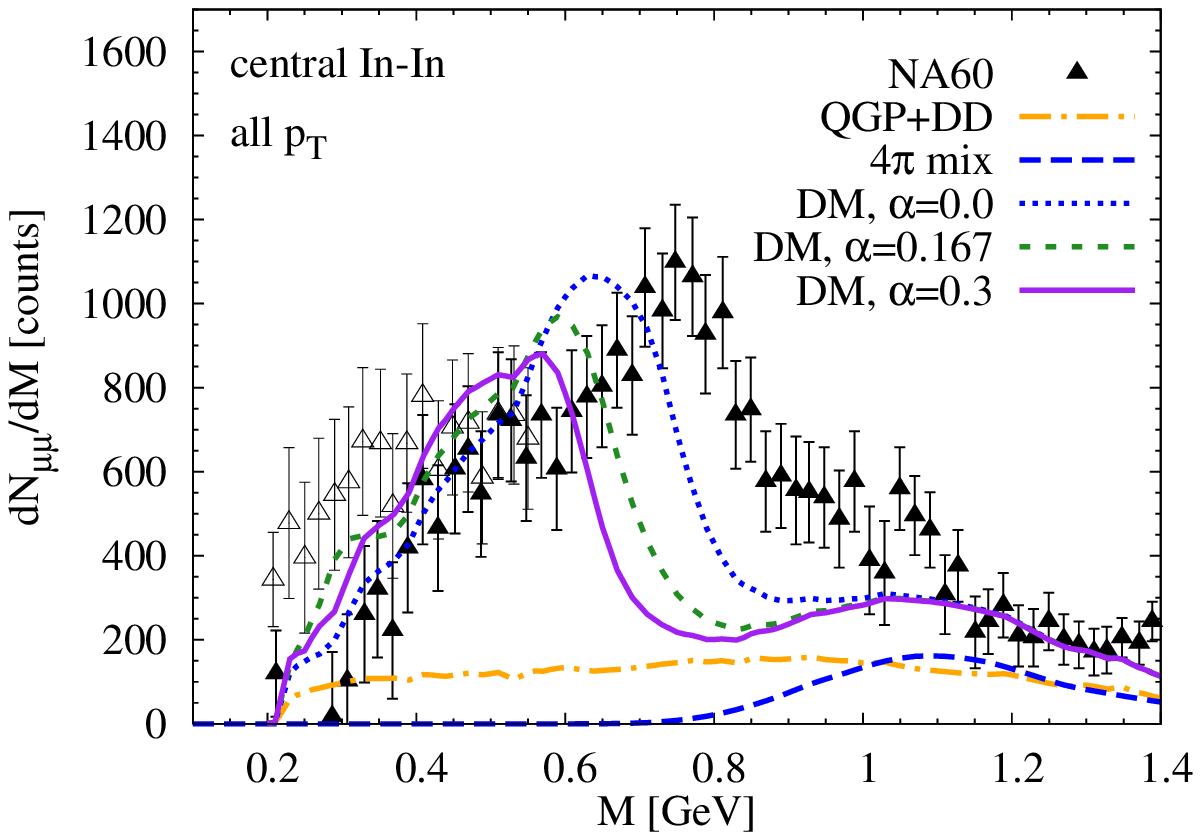}
\end{minipage}
\caption{\label{fig.3} NA60 data~\cite{dam05} compared to $\mu^+\mu^-$
  spectra from a thermal fireball using in-medium dilepton rates from
  the chiral-reduction formalism~\cite{syz97} (left panel) and
  from a dropping $\rho$-meson mass scenario (right panel). 
  The absolute normalization of the spectra, as following from the
  fireball, is the same as in Fig.~\ref{fig.2}.}
\end{figure}

The right panel of Fig.~\ref{fig.3} shows that the observed enhancement
below the free $\rho$ mass is accounted for, but the yield in the
peak region is underpredicted. Whether the introduction of a cocktail
component (free $\rho$) can provide a consistent resolution of this
discrepancy is questionable at present. Modifications of (or neglecting
altogether) the $T$-dependence of the dropping-mass parameterization
does not affect this conclusion.
 
Recent studies of axial-/vector-mesons within a renormalization-group
framework for the phase diagram of a generalized hidden-local-symmetry 
model at finite $T$ support the notion of dropping masses, but also 
imply a violation of the VDM~\cite{HS05}. Whether this approach can be
reconciled with the NA60 data is currently not known. An explicit 
calculation of pertinent spectral functions at finite temperature 
\emph{and} baryon-density is needed before further conclusions can be 
drawn.

%%%%%%%%%%%%%%%%%%%%%%%%%%%%%%%%%%%%
\section{Conclusions and Outlook}
%%%%%%%%%%%%%%%%%%%%%%%%%%%%%%%%%%%

Recent NA60 dimuon data have opened the possibility to study in-medium
modifications of vector mesons with unprecedented sensitivity. We have
confirmed that predictions from hadronic many-body approaches, whose
main feature is a strong broadening of the $\rho$ spectral function,
quantitatively describe the spectra in central In(158~$A$GeV)-In.  While
the absolute normalization is subject to current uncertainties of the
underlying fireball lifetime (which do not affect the spectral shape
significantly),
the relative strength of the different thermal sources is fixed. In the
mass region above 1~GeV, four-pion contributions to the e.m.~correlator
prevail and account for the observed enhancement, especially if effects
of chiral $V$-$A$ mixing are incorporated.  The NA60 dimuons thus
support a ``melting'' $\rho$-meson and the notion of a
``quark-hadron duality'' of the e.m.~spectral function around $T_c$. We
have also indicated that future analyses may be able to extract
information on in-medium $\omega$ and $\phi$ spectral functions, which
has not been possible to date.

We have investigated alternative approaches to describe the NA60 data. 
A naive dropping-mass scenario, as used to explain CERES data in the
1990's, is disfavored, unless a large component of free $\rho$ decays
can be argued for ($q_t$-spectra will provide valuable insights). The
chiral-reduction formalism is roughly in line with the spectra, but
quantitatively lacks $\rho$ broadening and/or low-mass enhancement.

To theoretically corroborate our findings, future studies within
chirally symmetric models are mandatory, including the effects of
baryons which we confirmed as important agents of medium effects.
Model-independent constraints, e.g.~from chiral~\cite{KS93} or QCD
sum rules~\cite{Ruppert05}, will enable further progress as well.

\section*{Acknowledgments}

We thank S.~Damjanovic and H.~Specht for information on the NA60
acceptance and many discussions. One of us (HvH) thanks the
A.-v.-Humboldt foundation for partial support via a Feodor-Lynen 
fellowship.  This work was supported in part by a U.S. National 
Science Foundation CAREER award under grant PHY-0449489.

\begin{flushleft}

\end{flushleft}

%\begin{flushleft}
%\bibliographystyle{lajolla}
%\bibliography{lajolla2006}
%\end{flushleft}

\end{document}